\documentclass[useAMS,usenatbib,usegraphicx]{mn2e}

\def \kms{\ifmmode{~{\rm km\,s}^{-1}}\else{~km~s$^{-1}$}\fi}
\def \vhel{\ifmmode{V_{{\rm hel}}}\else{$V_{{\rm hel}}$}\fi}
\def \vsys{\ifmmode{V_{{\rm sys}}}\else{$V_{{\rm sys}}$}\fi}
\def \degree{\ifmmode{^{\circ}}\else{$^{\circ}$}\fi}

\def \myr{\ifmmode{{\rm\ M}_\odot{\rm\ yr}^{-1}}\else{${\rm\ M}_\odot$ 
yr$^{-1}$}\fi}

\def \mdot{\ifmmode{{\rm\dot{M}}}\else{${\rm\dot{M}}$}\fi}
\def \msun{\ifmmode{{\rm\ M}_\odot}\else{${\rm\ M}_\odot$}\fi}
\def \rsun{\ifmmode{{\rm\ R}_\odot}\else{${\rm\ R}_\odot$}\fi}
\newcommand{\HA}{H$\alpha$}

\newcommand{\OIII}{[O\,{\sc iii}]\ $\lambda$5007\,\AA}

\newcommand{\HeII}{He\,{\sc ii}\ $\lambda$6560\,\AA}

\newcommand{\NII}{[N\,{\sc ii}]\ $\lambda$6584\,\AA}

\title[The morphology of Sp~1]{The morphology and kinematics of the Fine Ring Nebula, planetary nebula Sp~1, and the shaping influence of its binary central star\thanks{Based on the observations made with European Southern Observatory telescopes at the La Silla or Paranal Observatories under programme IDs 74.D-0373 and 55.D-0550.}}

\author[D. Jones et al.]{D. Jones$^{1}$\thanks{E-mail: djones@eso.org}, 
D. L. Mitchell$^{2,3}$, M.~Lloyd$^{3}$, 
D. Pollacco$^{4}$, T. J. O'Brien$^{3}$,\newauthor 
J. Meaburn$^{3}$ and N. M. H. Vaytet$^{5}$\\
$^{1}$European Southern Observatory, Alonso de C\'ordova 3107, Casilla 19001, Santiago, Chile\\
$^{2}$Chalmers Tekniska H\"ogskola, SE-412 96 Gothenburg, Sweden\\
$^{3}$Jodrell Bank Centre for Astrophysics, School of Physics and Astronomy, University of Manchester, M13 9PL, UK\\
$^{4}$Astrophysics Research Centre, Queen's University Belfast, BT7 1NN, UK\\
$^{5}$\'{E}cole Normale Sup\'{e}rieure de Lyon, CRAL (UMR CNRS 5574), Lyon, France
}

\begin{document}

\date{Accepted .  Received ; in original form }

\pagerange{\pageref{firstpage}--\pageref{lastpage}} \pubyear{}

\maketitle

\label{firstpage}

\begin{abstract}
We present the first detailed spatio-kinematical analysis and modelling of the planetary nebula Shapley 1 (Sp~1), which is known to contain a close-binary central star system.  Close-binary central stars have been identified as a likely source of shaping in planetary nebulae, but with little observational support to date.

Deep narrowband imaging in the light of \OIII\ suggests the presence of a large bow-shock to the west of the nebula, indicating that it is undergoing the first stages of an interaction with the interstellar medium.
Further narrowband imaging in the light of \HA{}+\NII\ combined with longslit observations of the \HA\ emission have been used to develop a spatio-kinematical model of Sp~1.  The model clearly reveals Sp~1 to be a bipolar, axisymmetric structure viewed almost pole-on.  The symmetry axis of the model nebula is within a few degrees of perpendicular to the orbital plane of the central binary system - strong evidence that the central close-binary system has played an important role in shaping the nebula.  

Sp~1 is one of very few nebulae to have this link, between nebular symmetry axis and binary plane, shown observationally. 
\end{abstract}


\section{Introduction}

Only a handful of planetary nebulae (PNe) are known to contain close-binary nuclei ($\sim$40, \citealt{demarco09}, \citealt{miszalski09a}, \citealt{miszalski11a}, \citealt{corradi11}, \citealt{miszalski11b}, \citealt{santander11}), yet it is widely believed that a close-binary central star is required to form an aspherical nebula (as the vast majority of PNe are).  The relatively small number of PNe known to host a central binary system is thought to be more a reflection of the difficulty in their detection than an accurate representation of the total number of central binary stars.

The generalised interacting stellar winds model (\citealp{kwok78}; \citealp{kahn85}; \citealp{balick02}) has proven the most effective theory in explaining some of the morphologies seen in PNe.  In this model, a slow, dense wind containing much of the envelope mass is blown at the end of the asymptotic giant branch (AGB) phase, and then swept up into a shell-like structure by a subsequent fast, tenuous wind from the emerging post-AGB star.  In order to recreate aspherical PNe, the slow AGB wind is assumed to be equatorially enhanced.  Only the effect of a close-binary partner is believed to provide the necessary equatorial enhancement to produce the most aspherical nebulae (magnetic fields have been considered, but cannot be sustained at the necessary levels for sufficient time to have the required effect, \citealp{nordhaus07}).  While theoretical support for this ``binary hypothesis" is abundant, observational evidence has been lacking.  In order to address this, the PLaN-B working group\footnote{http://www.wiyn.org/planb/} was formed to co-ordinate the investigative effort.

Spatio-kinematic modelling is an important tool in the testing of theoretical models, providing 3-D morphologies and orientations as well as the velocity field of the outflows and their kinematical ages, all of which must be replicated theoretically.  Additionally, for those PNe with well constrained binary systems, one can compare the parameters of the central binaries to those of their host nebulae, in particular the theoretical prediction that the nebular symmetry axis will lie perpendicular to the orbital plane of the binary system \citep{nordhaus06}.

\begin{figure*}
\centering
{\includegraphics[width=\textwidth]{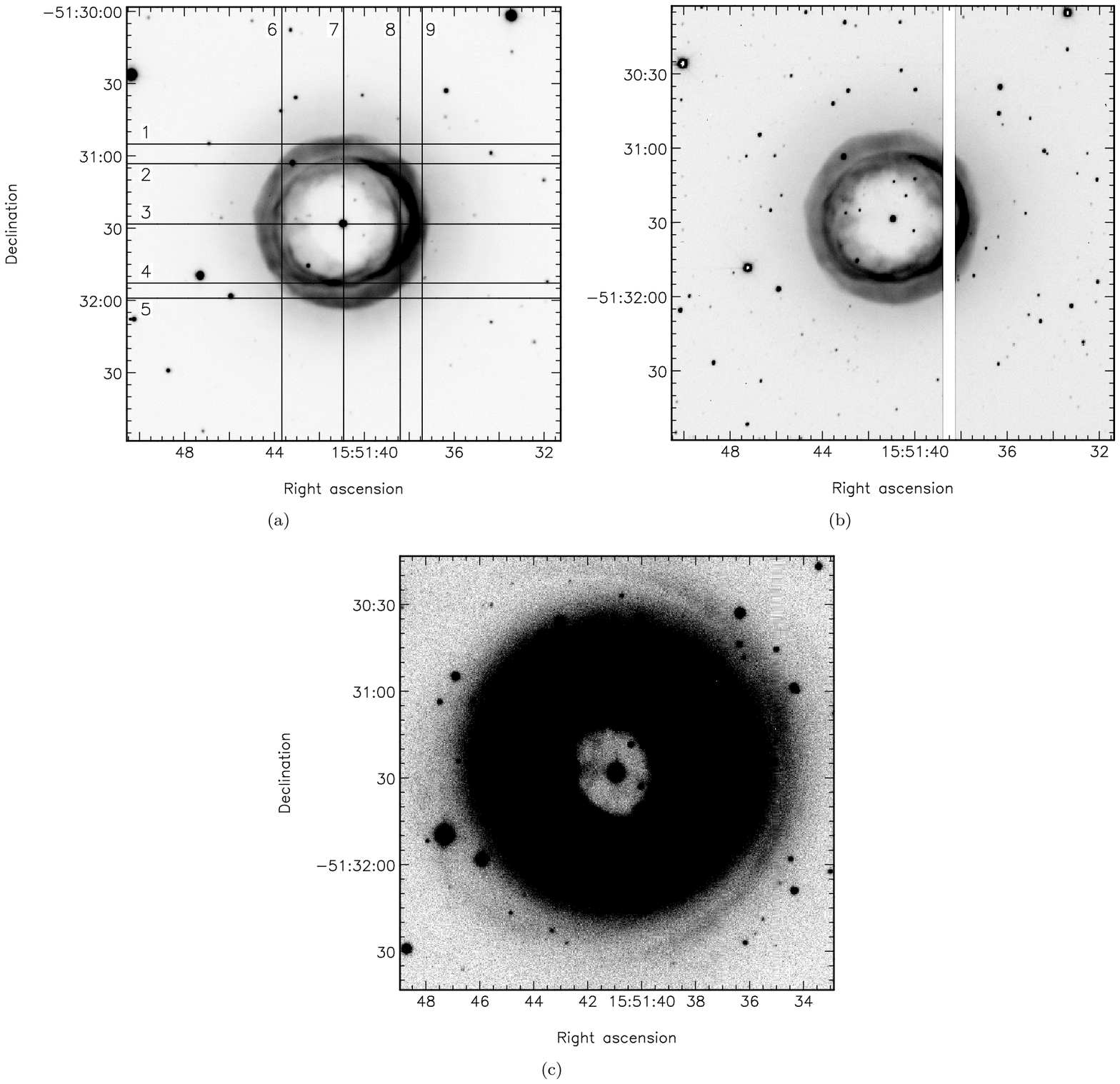}}
\caption[Deep imagery of Sp~1]{\label{Sp1_image}  
 Narrowband EMMI-NTT images of Sp~1 at  (a)\OIII\ (showing the observed slit positions) and (b) \HA{}. The vertical white stripe in the \HA\ image is a
result of the overscan gap between the master and slave CCD chips.
 The lower panel (c) shows the \OIII\ image but at low contrast to reveal the faint bowshock to the west of the main nebula.}
\end{figure*}

Sp~1 ($\alpha=15^{h}51^{m} 41^{s}$, $\delta=-51\degr 31' 23''$,
J2000), discovered by \citet{shapley36}, was described as ``nearly perfectly circular'' in appearance by \citeauthor{bond90} (\citeyear{bond90}, see Fig. \ref{Sp1_image}). This apparent morphology is somewhat at odds with presence of a binary central star, discovered spectroscopically by \citet{mendez88} and confirmed photometrically by \citet{bond90}, which would be expected to produce an aspherical nebula.  However, this can be reconciled by the hypothesis that Sp~1 is actually an axisymmetric nebula viewed almost pole-on, making Sp~1 morphologically akin to other PNe with known close-binary central stars (A~63: \citealp{mitchell07b}; A~41: \citealp{jones10b}; NGC6337: \citealp{garcia-diaz09}; HaTr~4: \citealp{tyndall11b}).  
This interpretation is `` supported by the low amplitude of the photometric variations of the central star and the absence of an eclipse'' \citep{bond90}.  Recent work by \citet{bodman11} and Hillwig et al. (in preparation) photometrically confirms the 2.9 day period of \citealt{bond90}, and indicates that the binary plane does indeed lie very close to the plane of the sky ($10\degr\geq i \geq 20\degr$\footnote{Here, the inclination, $i$, is defined such that for $i = 90$\degr\ the orbital plane would be in the line of sight (i.e. eclipsing).}).

Indeed, it has been proposed that many ring-like PNe may be bipolars with symmetry axes inclined to the plane of the sky; this suggestion is supported by kinematical studies of several ring-like PNe, all of which are found to be bipolar, e.g. the Ring Nebula \citep{bryce94}, LoTr~5 \citep{graham04}, the Helix Nebula \citep{meaburn05b}, SBW~1 \citep{smith07} and SuWt~2 \citep{jones10a}.

Despite the importance of Sp~1 as a test of current understanding of
how close-binaries affect the morphological evolution of PNe,
until now, there have been no kinematical observations of the nebula
in order to constrain its three dimensional structure. Detailed
imagery and high-resolution longslit spectroscopy are thus
presented in order to determine the true morphology of Sp~1.  The nebula morphology and orientation are then compared to that of the central binary in order to ascertain their relationship and test current theories of binary-induced PN shaping.

\section{Observations}
\label{sec:obs}
\subsection{NTT observations}
\label{sec:NTT}

Deep narrow-band \OIII\ and  \HA{}\footnote{The filter used also includes the \NII\ emission line, but the spectroscopy described later confirms that Sp~1 displays almost no \NII\ emission.} imagery of Sp~1 was obtained using the ESO (European Southern Observatory) Multi-Mode Instrument (EMMI; \citealp{dekker86}) on the 3.6-m ESO New Technology Telescope (NTT), and is shown in Fig. \ref{Sp1_image}.  The 1200-s \OIII\ image (Fig. \ref{Sp1_image}(a)) was taken on 1995 April 22 with a seeing of 1\arcsec\ and
shows the nebula as a diffuse ring in the
plane of the sky, upon which, bright filaments are superimposed. The
filaments give the appearance of two bright, narrow rings. It is not
clear whether they are two concentric rings or two rings that are
offset in the east-west direction. The filaments appear to intersect
in the south of the nebula, but there is no obvious counterpart in the
north, which would be expected if they are two offset rings. Both
rings are brighter in the west of the nebula than the east. Diffuse
material is present in the innermost region of the nebula and a very
faint outer ``halo'' is visible surrounding the bright nebular shell.
The 900-s \HA\ image (Fig. \ref{Sp1_image}(b)), taken on  2005 March 03 with a seeing of 0.9\arcsec{}, shows the same
bright ring-like filaments, however, the diffuse material visible
close to the central star appears brighter in \HA\ than in \OIII.

The \OIII\ image is also shown at low-contrast in Fig.~\ref{Sp1_image}c
to reveal, for the first time, a prominent bowshock to the west of
Sp~1.  The origin of this bowshock is discussed in Section \ref{sec:ISM}.

Spatially resolved, longslit emission-line spectra of Sp~1 were been obtained with EMMI on the NTT. Observations took place in 2005 March 2-4 using the red arm of the spectrograph which employs two MIT/LL CCDs, each of 2048 $\times$ 4096 15 $\mu$m pixels ($\equiv$ 0.166\arcsec\ per pixel), in a mosaic. There is a gap of 47 pixels ($\equiv 7.82\arcsec$ ) between the two CCD chips which can be seen in the observed spectra.
 
EMMI was used in single order echelle mode, with grating \#10 and a narrow-band \HA\ filter (\#596) to isolate the $87^{\rm{th}}$ echelle order containing the \HA\ and \NII\ emission lines.  Binning of \mbox{2 $\times$ 2} was used giving spatial and spectral scales of 0.33\arcsec\ per pixel and 3.8\kms\ per pixel, respectively.  The slit had a length of 330\arcsec{} and width 1\arcsec{} (\mbox{$\equiv$ 10 km~s$^{\rm -1}$}).  All integrations were of 1800~s duration and the seeing never exceeded 1\arcsec{}.

Data reduction was performed using \textsc{starlink} software. The spectra were bias-corrected and cleaned of cosmic rays. The spectra were then wavelength calibrated against a long exposure ThAr emission lamp, taken at the start of each night.  The calibration was confirmed using short Ne emission lamp exposures throughout the night, and by checking the wavelengths of skylines visible in the exposure. Finally the data were rescaled to a linear velocity scale (relative to the rest wavelength of \HA\ taken to be $6562.81$~\AA{}) and corrected for Heliocentric velocity.

\begin{figure*}
\centering
\includegraphics[]{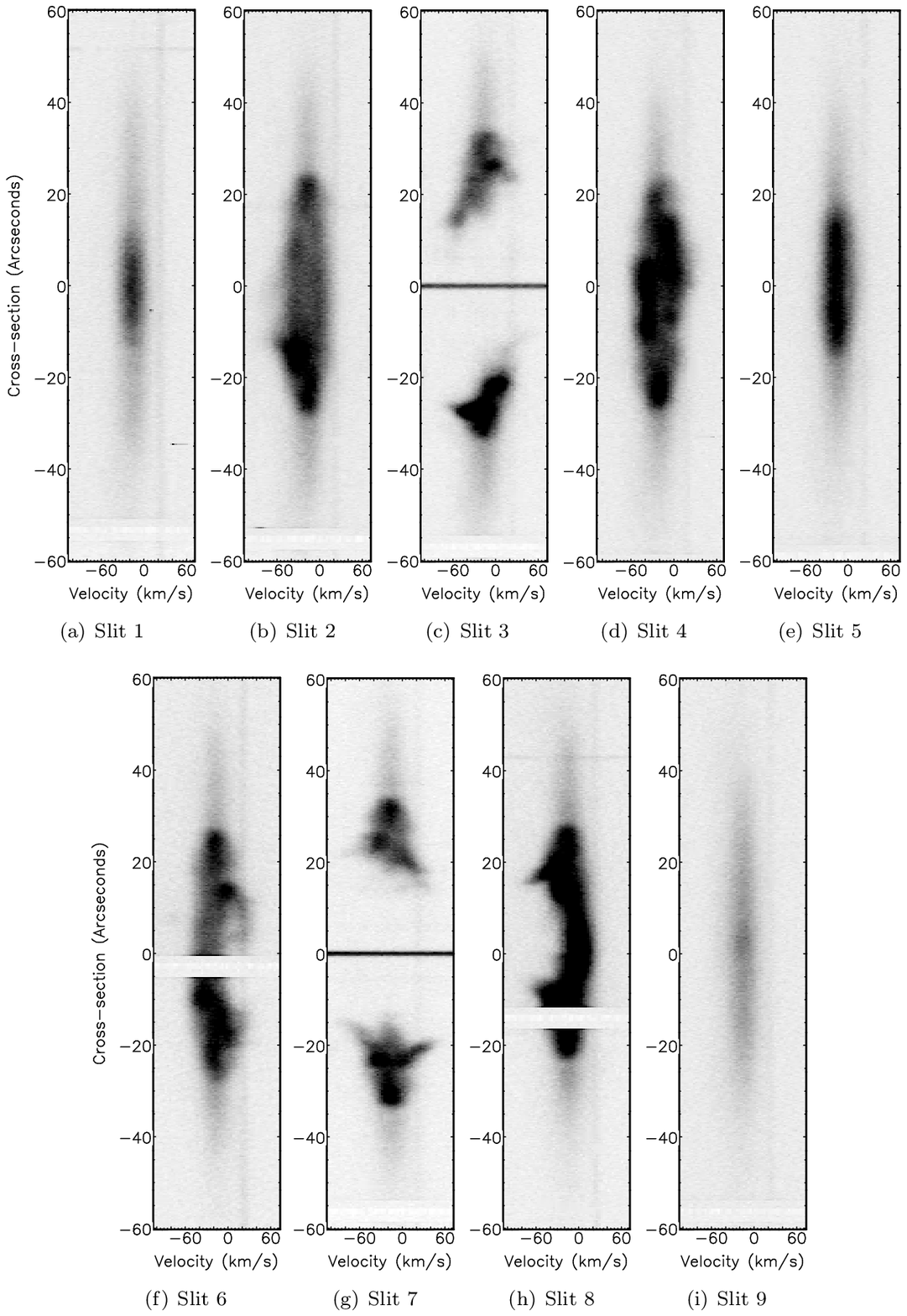}
\caption[\HA\ longslit spectra from Sp~1 taken with EMMI]{\label{NTT_spectra} \HA\ longslit spectra of Sp~1 obtained with EMMI combined with the NTT. The slit positions are drawn on Fig.~\ref{Sp1_image}(a). Cross-section zero corresponds to the position of the central star and the velocity axis on all plots is heliocentric velocity, $v_{hel}$. The horizontal white stripes that appear in some of the spectra result from the overscan gap between the master and slave CCD chips.}
\end{figure*}

Five slit positions were obtained with the slit orientated east-west
across Sp~1 (numbered 1 to 5) and four slit positions with the slit positioned in a
north-south direction (number 6 to 9). The slit positions are shown on
the \OIII\ image of Sp~1 in Fig.~\ref{Sp1_image}(a) (Note that the full 6\arcmin{} extent of the slits is not shown).  The fully reduced position-velocity (PV) arrays for \HA\ emission are shown in Fig.~\ref{NTT_spectra}, respectively (as there is very little \NII\ emission present, the spectra have been cropped to display \HA\ emission only).

\subsection{AAT spectroscopy}
\label{sec:AAT}
Complementary longslit observations of Sp~1 were obtained using UCLES combined with the 3.9-m f/36 AAT. UCLES was used in its primary spectral mode with a narrow-band filter to isolate the \HA{}+\NII\ emission lines in the
34th echelle order. Observations took place in January 2005 using an EEV2 CCD with 2048$\times$4096 13.5 $\mu$m square pixels, but using a 960$\times$4096 subsection of the CCD ($\equiv$ 0.16\arcsec pixel$^{-1}$). All integrations were of 1800-s duration.

Binning of 2$\times$3 was adopted during the observations, giving
2096 pixels in the spectral (x) direction ($\equiv$ 3.88 kms$^{-1}$
pixel$^{-1}$) and 320 pixels in the spatial (y) direction ($\equiv$
0.48\arcsec pixel$^{-1}$). This gave a projected slit length of 56.125\arcsec\ on the sky. The slit width was 1.971\arcsec\ ($\equiv$ 10 kms$^{-1}$).

Two integrations were obtained with the slit orientated north-south along the nebula, 5\arcsec\ either side of the central star.  Data reduction was performed in the same manner as for the EMMI spectra.  Unfortunately the observations were limited due to poor weather, and hence full cuts across the nebula were not obtained (single slits did not extend across the entire diameter of Sp~1) .  The data are consistent with the MES-SPM data shown here, and are available for download from The San Pedro M\'artir Kinematic Catalogue of Galactic Planetary Nebulae\footnote{http://kincatpn.astrosen.unam.mx} \citep{lopez12}.

\subsection{SAAO photometry}

Based on the period and amplitude of the reflection effect discussed by \citet{bond90}, we performed preliminary modelling of the central binary of Sp~1 using the Wilson-Devinney formulation \citep{wilson71}.  For a given primary component, the amplitude of the reflection effect is primarily governed by the size of the secondary component, the orbital period and inclination.  Assuming a low orbital inclination (as suspected from the nebular imagery, \citealp{bond90}), it was found that a secondary much larger than a main sequence M-dwarf is required to replicate the observations (just as found by \citealp{bodman11}).  This is similar to the results determined for the eclipsing central stars of Abell~46 and Abell~63, V477 Lyrae and UU Sge \citep{pollacco93,pollacco94}.  For orbital inclinations lower than $\sim$15\degr{}, we found that an M-dwarf secondary would need to be very close to filling its Roche lobe in order to replicate the reported period and amplitude of Sp~1.

High cadence photometry was acquired in order to look for signs of on-going mass transfer, from a Roche lobe filling secondary onto the primary, in the central star of Sp~1.  The central star of Sp~1 was observed in consecutive 90-s exposures in the I-band for between 1 and 3 hours on the nights of 2010 March 24, 25, 27, 28, 29, 31 and April 1, 2 and 6 using the SAAO CCD on the 1.9-m Radcliffe Telescope of the South African Astronomical Observatory (SAAO).  The SAAO CCD was employed with a SITe4 CCD binned $2\times2$ resulting in a pixel scale of 0.28\arcsec{} per pixel.  The seeing during the observations varied between 1\arcsec{} and 4\arcsec{}.  The resulting data were debiased and flat-fielded using standard \textsc{starlink} routines.

Differential photometry of the central star, with respect to a non-variable field star, was performed using \textsc{sextractor} \citep{sextractor} with photometric aperture of radius equal to 1.5$\times$ the seeing of each individual frame.  Due to the roughly 3 day period of the variability the resulting lightcurve is highly fragmentary (and as a result is not presented here), but entirely consistent with the 2.91 day period determined by \citet{bond90}.  No evidence for further short-timescale variability, which would be consistent with on-going accretion was found.  Further detailed study of the central star system (such as that of Hillwig et al.\ in prep.), including spectroscopic observations, is required to fully constrain the Roche lobe filling factor of the secondary.

\section{Analysis}
\label{sec:analysis}

\subsection{Longslit spectroscopy}
\label{sec:LSS}

\begin{figure}
\centering
\includegraphics[]{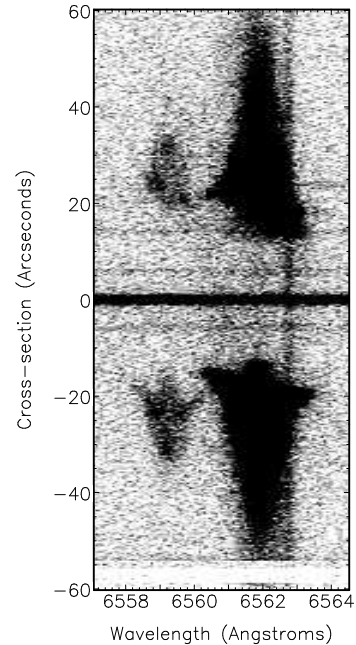}
\caption [Sp~1 longslit spectrum revealing \HeII\ emission]{\HA\ longslit spectrum (slit position~7) shown at low contrast in order to reveal the faint \HeII\ emission. The \HeII\ emission reflects the velocity structure of the \HA\ line. The \HA\ airglow line is present at 6562.817~\AA.}
\label{Sp1_HeII}
\end{figure}

All longslit spectra are bright in \HA\ and the nebular shell is also
faintly visible in \HeII\ emission. A longslit spectrum (from slit position
7) is shown at low contrast in Fig.~\ref{Sp1_HeII} to reveal the
\HeII\ emission. The \HeII\ emission emanates from the bright nebular
shell and ring-like filaments of Sp~1, but no \HeII\ emission is
visible from the faint, outer ``halo''. The presence of \HeII\ and absence
of \NII\ emission indicates that Sp~1 is a high excitation nebula with a very hot central star, entirely consistent with the temperature determined by \citet{frew08} of 72,000 Kelvin using the Zanstra Method.

The PV array from Slit 3 (Fig.\ \ref{NTT_spectra}(c)), which is orientated east-west (at a position angle, PA, of $-90$\degr{}) and crosses the central star of Sp~1, shows two bright emission regions corresponding to the east and west of the nebular shell.  These two regions are not joined in a velocity ellipse, indicating that Sp~1 is aspherical and open-ended.  Both emission regions show a ``chicken foot''-like profile with three spurs of emission directed towards the stellar continuum.  These three spurs can be attributed to the approaching lobe (blue-shifted spur), waist region (central spur) and receding lobe (red-shifted spur) of a bipolar shell.  Assuming an axisymmetric structure, the PV array also indicates that the nebular symmetry axis is probably inclined to the line of sight, as for an inclination of $0$\degr{} one would expect the PV array to be mirror symmetric about the central star.  As this is not the case, one can infer that the approaching lobe of the nebula is shifted to the east with respect to its receding counterpart (i.e. the blue-shifted spurs appear at more negative cross-sections than the corresponding red-shifted spurs).

The emission from Slit 7 (Fig.\ \ref{NTT_spectra}(g)), which runs north-south (at a PA of $0$\degr{}) and crosses the central star, displays a very similar ``chicken foot'' structure to the PV array from Slit 3 but without any obvious angular offset between blue- and red-shifted components.  This indicates that the small deviation of the nebular symmetry axis, from the line-of-sight, is almost entirely in the east-west direction.  This is confirmed by the PV arrays from Slits 2 and 4 ( Figs.\ \ref{NTT_spectra}(b) \& (d) crossing the north and south of the nebula, respectively, both at a PA of $-90$\degr{}), which both display velocity ellipses at roughly the same $V_{hel}$.

Slit positions 1 and 5 are positioned east-west across the large,
diffuse nebular shell and slit position 9 is orientated north-south
along the shell (see Fig.~\ref{Sp1_image}(a)). None of these slit
positions intersect with the bright nebular shell. The PV arrays from these Slits all show consistent velocity trends with a single velocity
component centred on $-18$ \kms{}.  All of the longslit spectra contain faint extended emission at this velocity, forming a filled ellipse in the PV arrays, indicative that this emission originates from a filled ``halo''  surrounding the brighter nebular shell rather than a hollow second shell.

\subsection{Interaction with the ISM}
\label{sec:ISM}

The presence of the bowshock in the deep \OIII\ image of Sp~1 indicates that it is undergoing an interaction with the surrounding interstellar medium (ISM), where the western side of the nebula is at the leading edge of its motion through the ISM.  However, as no proper motion measurements exist for this nebula it is not possible to confirm its direction of motion relative to that of the local ISM (assumed to be due to Galactic rotation, as in the analysis of HFG~1 - another PN with a binary central star - performed by \citealp{meaburn09}).  The bowshock places Sp~1 at stage 1 in the scheme of \citet{wareing07}, meaning that this interaction with the ISM cannot explain why the nebular shell appears brighter in the west than the east. Any material to the west of the bowshock is expected to be compressed and appear brighter, whilst material to the
east of the bowshock (including the nebular shell) will remain unperturbed \citep{wareing07}.

\section{The morpho-kinematics of Sp~1}
\subsection{Spatio-kinematical modelling}

A spatio-kinematical model, corresponding to the simplest three-dimensional structure consistent with the large-scale nebular \HA\ emission features, has been constructed for Sp~1.  The modelling was performed in order to confirm the axisymmetric nature of the nebula and to constrain the inclination of this symmetry axis for comparison with the inclination of the central binary.  The model was developed using {\sc novacart} \citep{gill99} and  assuming radial expansion (where the expansion velocity is proportional to distance from the nebular centre, commonly known as a Hubble-type flow).  The scale of the flow was set by the maximum observed velocity between red- and blue-shifted components in the nebular PV arrays (Fig. \ref{NTT_spectra}), measured to be 110\kms{}, indicating that the maximum expansion velocity at the tip of each lobe is $\sim$55\kms{}.  The systemic velocity of the nebula, $V_{sys}=-18$\kms{}, was set by the centroid of the faint extended emission attributed to the ``halo'' in Section \ref{sec:analysis}.  The other model parameters (e.g. dimensions and inclination) were manually varied over a wide range of values and the results compared by eye to both spectral observations and imagery, until a best fit was found.

\begin{figure*}
\centering
\includegraphics{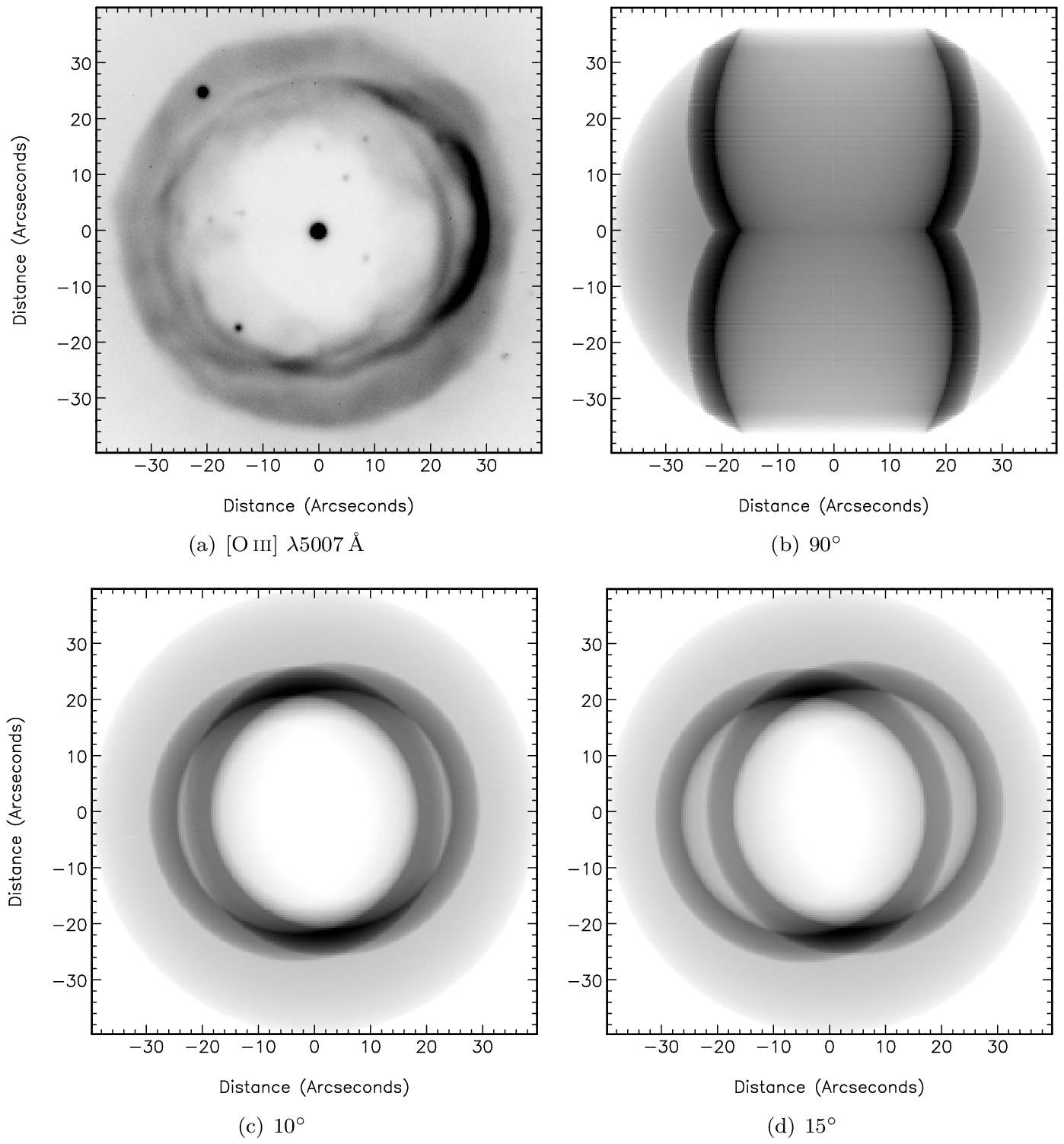}
\caption[Morphological-kinematical model of Sp~1 viewed at 90$^{\circ}$, 15$^{\circ}$ and 10$^{\circ}$]{\label{ellipse_model} The observed \OIII\ image of Sp~1 (a) alongside two-dimensional grey-scale images of the morphological-kinematical model viewed at (b) 90$^{\circ}$, (c) 10$^{\circ}$ and (d) 15$^{\circ}$.  The model nebula consists of a bright hourglass-shaped structure, embedded in a fainter ``halo'' (surrounding the waist of the nebula but not extending as far as the poles).}
\end{figure*}

The final model comprises an open-ended bipolar shell (with major to minor axis ratio of roughly 2:1) embedded in a filled spherical shell.  The
velocity of the filled, outer sphere was also proportional to radius
with a maximum velocity of 20\kms{}, and its density was set to
half that of the offset ellipsoids to replicate the brightness
contrast observed in Fig.~\ref{Sp1_image}. A two-dimensional
image of the model viewed in cross-section at an inclination of
90\degr is shown in Fig.~\ref{ellipse_model}a. Sp~1 was found to
have an inclination in the range 10-15\degr, and the model is
shown at inclinations 15\degr and 10\degr in
Figs.~\ref{ellipse_model}b and c, respectively.

The model successfully reproduces the two bright, offset rings and the larger diffuse ring when viewed at inclinations of
10$^{\circ}$ and 15$^{\circ}$. Both models also reproduce the faint
emission observed inside of the bright rings. The 10$^{\circ}$ model
reproduces the observed offset between the two rings most
convincingly.

The model is axisymmetric and therefore does not reproduce the
difference in brightness observed between the east and west sides of
Sp~1. The purpose of the model is simply to reproduce the large-scale velocity features
observed in the longslit spectra of Sp~1.

\begin{figure*}
\centering
\includegraphics{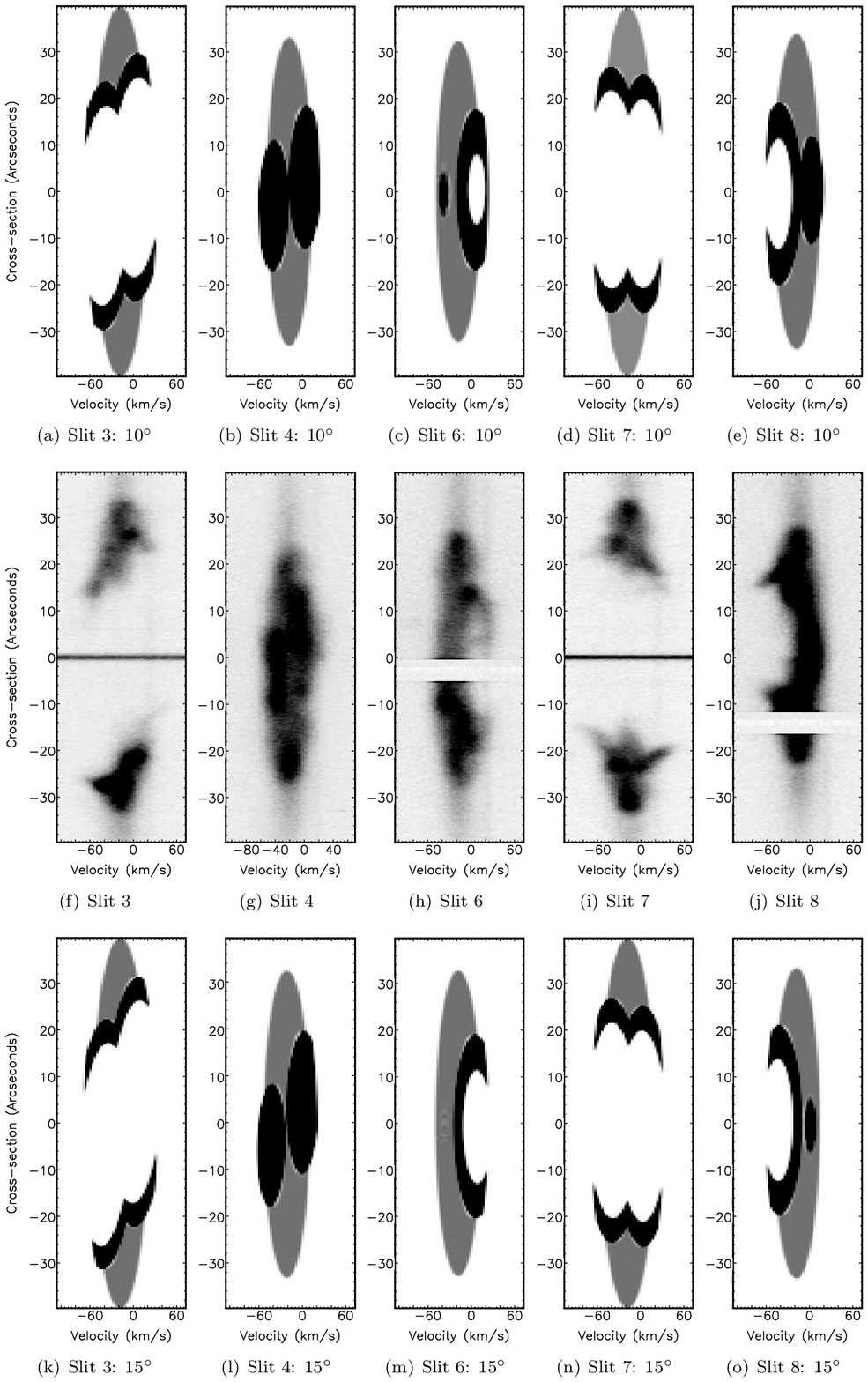}
\caption[Synthetic spectra extracted from morphological-kinematical model of Sp~1]{\label{modelcomp}Synthetic spectra extracted from the morphological-kinematical model of Sp~1 viewed at an inclination of 10$^{\circ}$ (top) and 15$^\circ$ (bottom), alongside the corresponding \HA\ longslit spectra shown at the same scale for comparison (middle).}
\end{figure*}

Synthetic spectra extracted from both the 10$^{\circ}$ and
15$^{\circ}$ models coinciding with slit positions 3, 4, 6, 7 and
8 are shown in Figs.~\ref{modelcomp}. Synthetic spectra from the other slit positions are not shown as the selected slits are sufficient to show the quality of fit provided by the model at the two inclinations.

The synthetic spectra extracted from the 10$^{\circ}$ model
(Fig.~\ref{modelcomp}) convincingly replicate the broad velocity
characteristics observed at all slit positions. The synthetic spectra
from positions 3 and 7 successfully reproduce the line splitting into
three velocity components as observed in the \HA\ longslit spectra;
however, the velocity of each component in the synthetic spectra does
not fan out to higher velocities as markedly as in the
observations.  The accurate reproduction of the central spur of emission from both positions 3 and 7 is clear evidence of the bipolar shell's pinched waist, as a more elliptical structure would only reproduce the red-shifted and blue-shifted spurs.

The synthetic spectrum from slit position 6 reproduces
the red-shifted velocity ellipse, which is present in the \HA\
spectrum. The synthetic spectrum extracted from slit position 8
successfully reproduces the incomplete blue-shifted velocity ellipse.

In general, the synthetic spectra extracted from the model viewed at
 15$^{\circ}$ (Fig.~\ref{modelcomp}) resemble the observed velocity
 trends in the NTT longslit spectra more closely than the 10$^{\circ}$
 model. At this inclination, the synthetic spectrum from slit position
 3 reflects the tilted structure observed in the \HA\ spectrum and the
 synthetic spectrum from slit position 4 demonstrates a greater offset
 between the bright red- and blue-shifted filaments, which is in close
 agreement with the \HA\ spectrum. Only the synthetic spectrum from
 slit position 6 fails to convincingly model the observed spectrum at
 an inclination of 15$^{\circ}$; the synthetic spectrum does not form
 a complete red-shifted velocity ellipse as seen in the NTT \HA\
 spectrum.
 
 Overall, both 10\degr{} and 15\degr{} models satisfactorily reproduce the observed images and spectra, with the 15\degr{} model providing a better fit to the spectroscopy and the 10\degr{} model a better fit to the imagery.  We, therefore, favour a nebular inclination between 10\degr{}--15\degr{}.  
 
 Comparison of both models with the observations shows no evidence for deviation from homologous expansion (Hubble-type flow) nor for any additional turbulent broadening in the nebular shell.  The modelling clearly proves that Sp~1 exhibits an hourglass-like morphology with a well-defined waist - no other morphology would be consistent with both the spectroscopy and imagery presented here.

\subsection{Kinematical age and systemic velocity}

The radial velocities produced by both the 10\degr\ and
15\degr\ spatio-kinematical models accurately reflect those
observed in the \HA\ longslit spectra. This suggests that the adopted
maximum expansion velocities of 55~\kms\ for the ellipsoids and
20~\kms\ for the faint, surrounding sphere are very good
approximations. The heliocentric systemic velocity of the model PN is $-18 \pm 5$ \kms{}, somewhat at odds with the value of $-31\pm3$\kms\ determined by \citet{meatheringham88}.  No reasonable explanation could be found for this discrepancy - spectra from slits 3 and 5, which cross the nebular centre, were collapsed and then fitted with a gaussian profile following the method of \citet{meatheringham88} determining a value consistent with that taken from the spatio-kinematical modelling.  To ensure that this was not an instrumental issue, the systemic velocity was confirmed using two UCLES spectra (Section \ref{sec:AAT}), again following the same method used by \citet{meatheringham88}.

The angular extent of the synthetic spectra is
consistent with the observed spectra, which suggests the adopted
geometrical dimensions in the model are also very good approximations. The
angular distance between the ends of both lobes is $\sim$70\arcsec\
(Fig.~\ref{ellipse_model}b). Adopting a distance to Sp~1 of 1.5~kpc
\citep{sabbadin86} gives a physical distance between the ends of both
lobes of 0.5 parsecs and a kinematical age of $\sim$8700 years.

\section{Discussion}

High spatial and spectral resolution long-slit \HA\ spectra have been obtained from the Fine Ring Nebula, Sp~1.  These spectra, together with deep narrow-band imagery, have been used to derive a spatio-kinematic model of the nebula proving its bipolar nature.  The spatio-kinematical model of Sp~1 fits with the classical
`butterfly' morphology for a PN defined in the classification scheme
of \citet{balick87}. The symmetry axis of the nebula is inclined almost along the line of sight ($10\degr\geq i \geq 15\degr$).  A Hubble-type flow is assumed with an equatorial expansion velocity of $\sim$25 \kms{}, consistent with typical PNe expansion velocities.  The heliocentric velocity of the PN was found to be $-18 \pm 5$ \kms{}.  The kinematical age of Sp~1, at a distance of 1.5 kpc \citep{sabbadin86}, is found to be $\sim$8700 years - well within the range of typical PN ages.

Sp~1 does \textit{not} exhibit an exceptional morphology amongst the known
sample of PNe with close-binary central stars.  The bipolar structure of the main nebula is fairly prevalent amongst other PNe with binary central stars that have also been subjected to detailed spatio-kinematical modelling \citep{jones11a,jones11b}.  The faint sphere of material which surrounds the bipolar shell, referred to in the text as a ``halo'', does not encompass the entire inner nebula, appearing mainly outside the waist of the bipolar shell.  It is, therefore, possible that the ``halo'' is more akin to an equatorial enhancement (such as the one in seen in HaTr~4, \citealp{tyndall11b}) than to the extended haloes found in some PNe \citep{corradi03}.  Equatorial enhancements have been identified as prevalent in the morphologies of PNe with close-binary central stars, which could be indicative that the formation of the ``halo'' in Sp~1 is in some way connected with the shedding of the CE (as is suspected for more markedly toroidal structures, \citealp{miszalski09b})

Sp~1 was found to be aligned approximately perpendicular to the plane of the central binary, consistent with theories of planetary nebula shaping by a close-binary central star.  This is one of very few nebulae to have had this link explicitly shown (along with A~63: \citealp{mitchell07b}, A~41: \citealp{jones10b}, NGC6337: \citealp{hillwig10}, A~65: \citealp{huckvale11} and HaTr~4: \citealp{tyndall11a,tyndall11b}), adding to the growing observational evidence that close-binary systems directly influence the morphological evolution of their host nebulae.

\section{Acknowledgments}
We thank the anonymous referee for their swift and useful comments.  We would also like thank Dr Alberto L\'opez, who obtained the AAT spectra and for useful discussions on the interpretation of the data. DJ thanks Dr. Todd Hillwig for several fruitful discussions about the central binary system of Sp~1, and Mar\'ia del Mar Rubio-D\'iez and Dr. Adam Avison for their assistance in the explication of the models.

We would like to thank the staff of the La Silla Paranal Observatory and the Anglo-Australian Telescope for their support in the acquisition of the spectroscopic observations.  This paper uses observations made at the South African Astronomical Observatory (SAAO).  
\bibliography{literature}

\begin{thebibliography}{27}
\expandafter\ifx\csname natexlab\endcsname\relax\def\natexlab#1{#1}\fi

\bibitem[{Balick(1987)}]{balick87}
Balick B., 1987, AJ, 94, 671

\bibitem[{Balick \& Frank(2002)}]{balick02}
Balick B., Frank A., 2002, A\&A Annual Review, 40, 439

\bibitem[\protect\citeauthoryear{{Bertin}}{{Bertin}}{1997}]{sextractor}
{Bertin} E.,  1997, {SExtractor User's Manual}.
Institut de Astrophysique \& Observatoire de Paris

\bibitem[{Bond \& Livio(1990)}]{bond90}
Bond H.E., Livio M., 1990, ApJ, 335, 568

\bibitem[{Bodman, Schaub \& Hillwig(2011)}]{bodman11}
Bodman E.~H.~L., Schaub S.~C., Hillwig T., 2011, Journal of the Southeastern Association for Research in Astronomy, in press

\bibitem[{Bryce et~al.(1994)Bryce, Balick \& Meaburn}]{bryce94}
Bryce M., Balick B., Meaburn J., 1994, MNRAS, 266, 721


\bibitem[{{Corradi} et~al.(2011){Corradi}, {Sabin}, {Miszalski},
  {Rodr{\'{\i}}guez-Gil}, {Santander-Garc{\'{\i}}a}, {Jones}, {Drew},
  {Mampaso}, {Barlow}, {Rubio-D{\'{\i}}ez}, {Casares}, {Viironen}, {Frew},
  {Giammanco}, {Greimel}, \& {Sale}}]{corradi11}
{Corradi}, R.~L.~M., {Sabin}, L., {Miszalski}, B., {Rodr{\'{\i}}guez-Gil}, P.,
  {Santander-Garc{\'{\i}}a}, M., {Jones}, D., {Drew}, J.~E., {Mampaso}, A.,
  {Barlow}, M.~J., {Rubio-D{\'{\i}}ez}, M.~M., {Casares}, J., {Viironen}, K.,
  {Frew}, D.~J., {Giammanco}, C., {Greimel}, R., \& {Sale}, S.~E. 2011, MNRAS,
  410, 1349
  
  \bibitem[{{Corradi} et~al.(2003){Corradi}, {Sch\"onberner}, {Steffen} \& {Perinotto}}]{corradi03}
Corradi R.L.M., Sch\"onberner D., Steffen M., Perinotto M., 2003, MNRAS, 340, 417

\bibitem[{{de Marco}(2009)}]{demarco09}
{de Marco} O., 2009, PASP, 121, 316

\bibitem[{{Dekker} et~al.(1986){Dekker}, {Delabre} \& {Dodorico}}]{dekker86}
{Dekker} H., {Delabre} B., {Dodorico} S., 1986, In {Crawford} D.L., ed.,
  Society of Photo-Optical Instrumentation Engineers (SPIE) Conference Series,
  vol. 627 of Society of Photo-Optical Instrumentation Engineers (SPIE)
  Conference Series, pp. 339--348

\bibitem[{{Frew}(2008)}]{frew08}
{Frew} D., 2008, PhD Thesis, Macquarie University

\bibitem[{{Garc{\'{\i}}a-D{\'{\i}}az} et~al.(2009){Garc{\'{\i}}a-D{\'{\i}}az},
  {Clark}, {L{\'o}pez}, {Steffen} \& {Richer}}]{garcia-diaz09}
{Garc{\'{\i}}a-D{\'{\i}}az} M.T., {Clark} D.M., {L{\'o}pez} J.A., {Steffen} W.,
  {Richer} M.G., 2009, ApJ, 699, 1633

\bibitem[{Gill \& O'Brien(1999)}]{gill99}
Gill C.D., O'Brien T.J., 1999, MNRAS, 307, 677

\bibitem[{{Graham} et~al.(2004){Graham}, {Meaburn}, {L{\'o}pez}, {Harman} \&
  {Holloway}}]{graham04}
{Graham} M.F., {Meaburn} J., {L{\'o}pez} J.A., {Harman} D.J., {Holloway} A.J.,
  2004, MNRAS, 347, 1370

\bibitem[{{Huckvale} et~al.(2011){Huckvale}, {Prouse}, {Jones}, {Lloyd} \& {Pollacco}}]{huckvale11}
{Huckvale} L., {Prouse} B., {Jones} D., {Lloyd} M., {Pollacco} D., 2011, In Asymmetric Planetary Nebulae 5 Conference, p.~113

\bibitem[{{Hillwig} et~al.(2010){Hillwig}, {Bond}, {Af{\c s}ar} \& {De
  Marco}}]{hillwig10}
{Hillwig} T.C., {Bond} H.E., {Af{\c s}ar} M., {De Marco} O., 2010, AJ, 140, 319

\bibitem[{{Jones} et~al.(2010{\natexlab{a}}){Jones}, {Lloyd}, {Mitchell},
  {Pollacco}, {O'Brien} \& {Vaytet}}]{jones10a}
{Jones} D., {Lloyd} M., {Mitchell} D.L., {Pollacco} D.L., {O'Brien} T.J.,
  {Vaytet} N.M.H., 2010{\natexlab{a}}, MNRAS, 401, 405

\bibitem[{{Jones} et~al.(2010{\natexlab{b}})}]{jones10b}
{Jones} D., {Lloyd} M., {Santander-Garc\'ia} M., {L\'opez} J.A., {Meaburn} J., {Mitchell} D.L., {O'Brien} T.J., {Pollacco} D., {Rubio-D\'iez} M.M., {Vaytet} N.M.H., 2010{\natexlab{b}}, MNRAS, 408, 2312

\bibitem[{{Jones} et~al.(2011a){Jones}, {Tyndall}, {Huckvale}, {Prouse} \& {Lloyd}}]{jones11a}
{Jones} D., {Tyndall} A.A., {Huckvale} L., {Prouse} B., {Lloyd} M., 2011a, In Schmidtobreick L., Schreiber M.R., Tappert C., eds, ASP Conf. Ser. Vol. 447, Evolution of Compact Binaries. Astron. Soc. Pac., San Francisco, p.~165

\bibitem[{{Jones} et~al.(2011b){Jones}, {Tyndall}, {Lloyd} \& {Santander-Gar\'ia}}]{jones11b}
{Jones} D., {Tyndall} A.A., {Lloyd} M., {Santander-Garc\'ia} M., 2011b, In IAUS 283: Planetary Nebulae: an Eye to the Future, in press

\bibitem[{Kahn \& West(1985)}]{kahn85}
Kahn F.D., West K.A., 1985, MNRAS, 212, 837

\bibitem[{Kwok et~al.(1978)Kwok, Purton \& Fitzgerald}]{kwok78}
Kwok S., Purton C.R., Fitzgerald P.M., 1978, ApJ, 219, 125

\bibitem[{L\'opez et~al.(2012)L\'opez, Richer, Garc\'ia-D\'iaz, Clark, Meaburn, Riesgo, Steffen \& Lloyd}]{lopez12}
L\'opez J. A., Richer M. G., Garc\'ia-D\'iaz Ma. T., Clark D. M., Meaburn J., Riesgo H., Steffen W., Lloyd M.,
2012, RevMexAA, 48, 3

\bibitem[{{Meaburn} et~al.(2005){Meaburn}, {Boumis}, {L{\'o}pez}, {Harman},
  {Bryce}, {Redman} \& {Mavromatakis}}]{meaburn05b}
{Meaburn} J., {Boumis} P., {L{\'o}pez} J.A., {Harman} D.J., {Bryce} M.,
  {Redman} M.P., {Mavromatakis} F., 2005, MNRAS, 360, 963

\bibitem[{{Meaburn} et~al.(2009){Meaburn}, {L{\'o}pez}, {Boumis}, {Lloyd} \&
  {Redman}}]{meaburn09}
{Meaburn} J., {L{\'o}pez} J.A., {Boumis} P., {Lloyd} M., {Redman} M.P., 2009,
  A\&A, 500, 827
  
  \bibitem[{{Meatheringham} et~al.(1988){Meatheringham}, {Wood}, \&
  {Faulkner}}]{meatheringham88}
{Meatheringham} S.J., {Wood} P.R., {Faulkner} D.J.., 1988,
  ApJ, 334, 862

\bibitem[{M\'endez et~al.(1988)M\'endez, Kudritzki, Herrero, Husfeld \&
  Groth}]{mendez88}
M\'endez R.H., Kudritzki R.P., Herrero A., Husfeld D., Groth H.G., 1988, A\&A,
  190, 113

\bibitem[{{Miszalski} et~al.(2009{\natexlab{a}}){Miszalski}, {Acker}, {Moffat},
  {Parker} \& {Udalski}}]{miszalski09a}
{Miszalski} B., {Acker} A., {Moffat} A.F.J., {Parker} Q.A., {Udalski} A.,
  2009{\natexlab{a}}, A\&A, 496, 813

\bibitem[{{Miszalski} et~al.(2009{\natexlab{b}}){Miszalski}, {Acker}, {Parker}
  \& {Moffat}}]{miszalski09b}
{Miszalski} B., {Acker} A., {Parker} Q.A., {Moffat} A.F.J., 2009{\natexlab{b}},
  A\&A, 505, 249
  
  \bibitem[{{Miszalski} et~al.(2011{\natexlab{a}}){Miszalski}, {Corradi}, {Boffin}, 
	{Jones}, {Sabin}, {Santander-Garc{\'{\i}}a}, 
	{Rodr{\'{\i}}guez-Gil} \& {Rubio-D{\'{\i}}ez}}]{miszalski11a}
{Miszalski} B., {Corradi} R.~L.~M., {Boffin} H.~M.~J., 
	{Jones} D., {Sabin} L., {Santander-Garc{\'{\i}}a} M., 
	{Rodr{\'{\i}}guez-Gil} P., {Rubio-D{\'{\i}}ez} M.~M., 2011{\natexlab{a}},
MNRAS, 413, 1264
  
  \bibitem[{{Miszalski} et~al.(2011{\natexlab{b}}){Miszalski}, {Jones}, {Rodr{\'{\i}}guez-Gil}, {Boffin}, {Corradi} \& {Santander-Garc{\'{\i}}a}}]{miszalski11b}
{Miszalski} B., {Jones} D., {Rodr{\'{\i}}guez-Gil} P., 
	{Boffin} H.~M.~J., {Corradi} R.~L.~M., {Santander-Garc{\'{\i}}a} M., 2011{\natexlab{b}},
  A\&A, 531, 158

\bibitem[{Mitchell et~al.(2007)Mitchell, Pollacco, O'Brien, Bryce, Lopez,
  Meaburn \& Vaytet}]{mitchell07b}
Mitchell D.L., Pollacco D., O'Brien T.J., Bryce M., Lopez J.A., Meaburn J.,
  Vaytet N.M.H., 2007, MNRAS, 374, 1404

\bibitem[{Nordhaus \& Blackman(2006)}]{nordhaus06}
Nordhaus J., Blackman E.G., 2006, MNRAS, 370, 2004

\bibitem[{{Nordhaus} et~al.(2007){Nordhaus}, {Blackman} \&
  {Frank}}]{nordhaus07}
{Nordhaus} J., {Blackman} E.G., {Frank} A., 2007, MNRAS, 376, 599

\bibitem[{Pollacco \& Bell(1993)}]{pollacco93}
Pollacco D.L., Bell S.A., 1993, MNRAS, 262, 377

\bibitem[{Pollacco \& Bell(1994)}]{pollacco94}
Pollacco D.L., Bell S.A., 1994, MNRAS, 267, 452

\bibitem[{{Sabbadin}(1986)}]{sabbadin86}
{Sabbadin} F., 1986, A\&AS, 64, 579

\bibitem[{{Santander-Garc\'ia} et~al.(2011){Santander-Garc\'ia}, {Rodr{\'{\i}}guez-Gil}, {Jones}, 
	{Corradi}, {Miszalski}, {Pyrzas} \&
	{Rubio-D{\'{\i}}ez}}]{santander11}
{Santander-Garc\'ia} M., {Rodr{\'{\i}}guez-Gil} P., {Jones} D., 
	{Corradi} R.~L.~M., {Miszalski} B., {Pyrzas} S., 
	{Rubio-D{\'{\i}}ez} M.~M., 2011, in {Asymmetric Planetary Nebulae 5}, edited by
  {{Zijlstra}, A.~A. and {McDonald}, I. and {Lagadec}, E.}, {Asymmetric
  Planetary Nebulae}, 259

\bibitem[{{Shapley}(1936)}]{shapley36}
{Shapley} H., 1936, Harvard College Observatory Bulletin, 902, 26

\bibitem[{Smith et~al.(2007)Smith, Bally \& Walawender}]{smith07}
Smith N., Bally J., Walawender J., 2007, AJ, 134, 846

\bibitem[{{Tyndall} et~al.(2011a){Tyndall}, {Jones}, {Lloyd}, {O'Brien}, {Pollacco} \& {Mitchell}}]{tyndall11a}
{Tyndall} A.A., {Jones} D., {Lloyd} M., {O'Brien} T.J., {Pollacco} D.L., {Mitchell} D.L., 2011a, In Asymmetric Planetary Nebulae 5 Conference, p.~115

\bibitem[{{Tyndall} et~al.(2011b){Tyndall}, {Jones} \& {Lloyd}}]{tyndall11b}
{Tyndall} A.A., {Jones} D., {Lloyd} M., 2011b, In IAUS 283: Planetary Nebulae: an Eye to the Future, in press

\bibitem[{{Wareing} et~al.(2007){Wareing}, {Zijlstra} \& {O'Brien}}]{wareing07}
{Wareing} C.J., {Zijlstra} A.A., {O'Brien} T.J., 2007, MNRAS, 382, 1233

\bibitem[{{Wilson} \& {Devinney}(1971){Wilson} \& {Devinney}}]{wilson71}
{Wilson} R.E., {Devinney} E.J., 1971, ApJ, 166, 605

\end{thebibliography}
\bibliographystyle{mnras}

\end{document}